\documentclass[conference,letterpaper]{IEEEtran}
\IEEEoverridecommandlockouts

\usepackage{tikz}
\usepackage{textcomp}
\usepackage{hyperref}
\usepackage{lipsum}

\newcommand\copyrighttext{%
  \footnotesize \textcopyright 2016 IEEE. Personal use of this material is permitted.
  Permission from IEEE must be obtained for all other uses, in any current or future
  media, including reprinting/republishing this material for advertising or promotional
  purposes, creating new collective works, for resale or redistribution to servers or
  lists, or reuse of any copyrighted component of this work in other works.
  DOI: \href{<http://tex.stackexchange.com>}{<DOI No.>}}
\newcommand\copyrightnotice{%
\begin{tikzpicture}[remember picture,overlay]
\node[anchor=south,yshift=10pt] at (current page.south) {\fbox{\parbox{\dimexpr\textwidth-\fboxsep-\fboxrule\relax}{\copyrighttext}}};
\end{tikzpicture}%
}
\usepackage{graphicx}
\usepackage{subcaption}
\usepackage{fixltx2e}
\usepackage{gensymb}
\usepackage{todonotes}
\usepackage{url}
\begin{document}
\title{The Matsu Wheel: A Cloud-based Framework 
\protect\\ for the Efficient Analysis and Reanalysis
\protect\\ of Earth Satellite Imagery}

\author{\IEEEauthorblockN{Maria T. Patterson\IEEEauthorrefmark{1},
Nikolas Anderson\IEEEauthorrefmark{1}, 
Collin Bennett\IEEEauthorrefmark{2},
Jacob Bruggemann\IEEEauthorrefmark{1}, 
Robert L. Grossman\IEEEauthorrefmark{1}\IEEEauthorrefmark{2},\\
Matthew Handy\IEEEauthorrefmark{3},
Vuong Ly\IEEEauthorrefmark{3},
Daniel J. Mandl\IEEEauthorrefmark{3},
Shane Pederson\IEEEauthorrefmark{2},
James Pivarski\IEEEauthorrefmark{2},\\
Ray Powell\IEEEauthorrefmark{1}, 
Jonathan Spring\IEEEauthorrefmark{1},
Walt Wells\IEEEauthorrefmark{4}, and
John Xia\IEEEauthorrefmark{1}}
\IEEEauthorblockA{\IEEEauthorrefmark{1}Center for Data Intensive Science\\
University of Chicago,
Chicago, IL 60637\\ mtpatter@uchicago.edu}
\IEEEauthorblockA{\IEEEauthorrefmark{2}Open Data Group\\
River Forest, IL 60305}
\IEEEauthorblockA{\IEEEauthorrefmark{3}NASA Goddard Space Flight Center\\
Greenbelt, MD 20771}
\IEEEauthorblockA{\IEEEauthorrefmark{4}Open Commons Consortium}
}

\maketitle
\thispagestyle{plain}
\pagestyle{plain}
\copyrightnotice

\begin{abstract}
Project Matsu is a collaboration between the Open Commons Consortium and NASA focused on developing open source technology for the cloud-based processing of Earth satellite imagery.  A particular focus is the development of applications for detecting fires and floods to help support natural disaster detection and relief. Project Matsu has developed an open source cloud-based infrastructure to process, analyze, and reanalyze large collections of hyperspectral satellite image data using OpenStack, Hadoop, MapReduce, Storm and related technologies. 

We describe a framework for efficient analysis of large amounts of data called the Matsu ``Wheel."  The Matsu Wheel is currently used to process incoming hyperspectral satellite data produced daily by NASA's Earth Observing-1 (EO-1) satellite. The framework is designed to be able to support scanning queries using cloud computing applications, such as Hadoop and Accumulo.  A scanning query processes all, or most of the data, in a database or data repository.  %

We also describe our preliminary Wheel analytics, including an anomaly detector for rare spectral signatures or thermal anomalies in hyperspectral data and a land cover classifier that can be used for water and flood detection. Each of these analytics can generate visual reports accessible via the web for the public and interested decision makers. The resultant products of the analytics are also made accessible through an Open Geospatial Compliant (OGC)-compliant Web Map Service (WMS) for further distribution.  The Matsu Wheel allows many shared data services to be performed together to efficiently use resources for processing hyperspectral satellite image data and other, e.g., large environmental datasets that may be analyzed for many purposes.

\end{abstract}
\IEEEpeerreviewmaketitle

\section{Introduction}
The increasing availability of large volumes of scientific data due to the
decreasing cost of storage and processing power has led to new challenges in
scientific research.  Scientists are finding that the bottleneck to 
discovery is no longer a lack of data but an inability to manage and 
analyze their large datasets. 

A common class of problems require applying an analytic computation over an 
entire dataset.  Sometimes these are called {\em scanning queries} since they involve a
scan of the entire dataset. For example, analyzing each image in a large collection
of images is an example of a scanning query.  In contrast, standard queries typically process a relatively small percentage of the data in a database or data repository.

With multiple scanning queries that are run within a time that is comparable to 
the length of time required for a single scan, it can be much more efficient to 
scan the entire dataset once and apply each analytic in turn versus scanning the
entire dataset for each scanning query as the query arrives. This is the case unless the data management infrastructure has specific technology for recognizing and processing scanning queries.  In this paper, we introduce a software application called the Matsu Wheel that is designed to support multiple scanning queries over satellite imagery data.

Project Matsu is a collaborative effort between the 
Open Science Data Cloud (OSDC), managed by the Open Commons Consortium (OCC), 
and NASA, working to develop open source
tools for processing and analyzing Earth satellite imagery in the cloud.
The Project Matsu ``Wheel" is a framework for simplifying Earth satellite image
analysis on large volumes of data by providing an efficient system that performs all 
the common data services and then passes the prepared chunks of data in a common format to the analytics, which processes each new chunk of data in turn.

\subsection{Motivation for an analytic wheel}
The idea behind the wheel is to have all of the data processing services 
performed together on chunks of data to efficiently use resources, including 
available network bandwidth, access to secondary storage, and available computing resources.  This is especially important with reanalysis, in which the entire dataset is processed using an updated algorithm, a recalibration of the data, a new normalization of the data, a new workflow, etc.   
The motivation behind the analytic wheel is 
to streamline and share these services so that they are only performed 
once on each chunk of data in a scanning query, regardless of the number or type of scanning analytics run over the data.

\subsection{Comparison to existing frameworks}
The Matsu Wheel approach differs from other common data management or data processing systems.  Real-time distributed data processing frameworks (for example, Storm, S4, or Akka) are designed to process data in real time as it flows through a distributed system (see also, e.g., \cite{Backman2012,Zeng2013,Kienzler2012}).  In contrast, the Wheel is designed for the reanalysis of an entire static dataset that is stored in distributed storage system (for example, the Hadoop Distributed File System) or distributed database (for example, HBase or Accumulo).

It is important to note, that any distributed scale-out data processing system based upon virtual machines has certain performance issues due to the shared workload across multiple virtual machines associated with a single physical node. In particular, these types of applications may have significant variability in performance for real scientific workloads \cite{Jackson2010}.  This is true, when multiple scanning queries hit a distributed file system, a NoSQL database, or a wheel based system on top of one these infrastructures.
When a NoSQL database is used for multiple scanning queries with a framework like the wheel, the NoSQL database can quickly become overloaded.

\subsection{Application to Earth satellite data}

Analyses of Earth satellite data and hyperspectral imagery data in particular
benefit from the Matsu Wheel system as a use case in which the data may be large,
have high-volume throughput, and are used for many types of applications.
The Project Matsu Wheel currently processes the data produced each day by NASA's 
Earth Observing- 1 (EO-1) satellite and makes a variety of data products available 
to the community.  In addition to the Atmospheric Corrector, the EO-1 satellite has 
two primary scientific instruments for land
observations, the Advanced Land Imager (ALI) and a hyperspectral 
imager called Hyperion \cite{Hearn2001,Pearlman2001}.  
EO-1 was launched in November 2000 as
part of NASA's New Millennium Program (NMP) initiative for advancing new technologies
in space and is currently in an extended mission.  

The ALI instrument acquires
data in 9 different bands from 0.48$-$2.35 $\mu$m with 30-meter resolution plus
a panchromatic band with higher 10-meter spatial resolution.  The standard scene
size projected on the Earth surface equates to 37 km x 42 km (width x length).
Hyperion has similar spatial resolution but higher spectral resolution, observing in 
242 band channels from 0.357$-$2.576 $\mu$m with 10-nm bandwidth. Hyperion scenes
have a smaller standard footprint width of 7.7 km.  The Matsu Wheel runs 
analytics over Level 1G data in Geographic Tagged 
Image File Format (GeoTiff) format, which have been radiometrically
corrected, resampled for geometric correction, and registered to a geographic
map projection.  The GeoTiff data and metadata for all bands in a single Hyperion
scene can amount to 1.5$-$2.5 GB of data for only the Level 1G data.  
A cloud environment for shared storage and computing capabilities is ideal for
scientific analysis of many scenes, which can quickly add to a large amount of data.

\section{Workflow}

\subsection{Cloud environment}

Project Matsu uses both an OpenStack-based computing platform and a 
Hadoop-based computing platform, both of which are managed by the OCC (www.occ-data.org)
in conjunction with the University of Chicago. The OpenStack platform 
(the Open Science Data Cloud \cite{Grossman2012}) currently contains 60 nodes,
1208 compute cores, 4832 GB of compute RAM, and 1096 TB of raw storage. 
The Hadoop \cite{White2012} platform currently contains 28 nodes, 896 compute cores, 
261 TB of storage, and 3584 GB of compute RAM. 

\subsection{Pre-processing of data on the OSDC}

The Open Science Data Cloud provides a number of data center services for 
Project Matsu.   
The data are received daily from NASA, stored on a distributed, fault-tolerant
file system (GlusterFS), and pre-processed prior to the application of the Wheel 
analytics on Skidmore.  The images are converted into SequenceFile format,
a file format more suited for MapReduce, and uploaded into HDFS \cite{MapReduce}.
Metadata and compute summary statistics are extracted for each scene and stored in 
Accumulo, a distributed NoSQL database \cite{Accumulo}.  The metadata are
used to display the geospatial location of scenes via a mapping service
so that users can easily visualize which areas of the Earth are covered in the
data processed by the Matsu Wheel.

Here is an overview of the Matsu data flow for processing EO-1 images and 
producing data and analytic products:
\begin{enumerate} 
\item Performed by NASA/GSFC as part of their daily operations: \\
a) Transmit data from NASA's EO-1 Satellite to NASA ground stations and then to
NASA/GSFC.\\
 b) Align data and generate Level 0 images.\\
 c) Transmit Level 0 data from NASA/GSFC to the OSDC.
\item Run by NASA on the OSDC OpenStack cloud for Matsu and other projects:\\
 a) Store Level 0 images in the OSDC Public Data Commons for long-term, active
storage.\\
 b) Within the OSDC, launch Virtual Machines (VMs) specifically built to 
 render Level 1 images from Level 0. Each Level 1 band is saved as a distinct 
 image file (GeoTIFF).\\
 c) Store Level 1 band images in the OSDC Public Data Commons for long-term storage.  \item Run specifically for Project Matsu on the Hadoop cloud:\\
 a) Read Level 1 images, combine bands, and serialize image bands into a single file.\\
 b) Store serialized files on HDFS.\\
 c) Run Wheel analytics on the serialized Level 1 images stored in HDFS.\\
 d) Store the results of the analysis in Accumulo for further analysis, generate reports, and load into a Web Map Service. 
\end{enumerate}

\subsection{Analytic `wheel' architecture}

The analytic Wheel is so named because multiple analytics are applied to data 
as it flows underneath.  While a bicycle or water wheel does not fit exactly, the
image is clear: do the work as the data flows through once, like the pavement under
the bicycle wheel.  With big data, retrieving or processing data multiple times
is an inefficient use of resources and should be avoided.  

When new data become available in HDFS as part of the pre-processing described
above, the MapReduce scanning analytics kick off.  
Intermediate output is written to HDFS, and all analytic results are stored in
Accumulo as JSON.  Secondary analysis that can run from the results of other
analytics can be done ``off the wheel" by using the Accumulo-stored JSON as
input.

\begin{figure*}[h ] 
\centering
\includegraphics[scale = .55]{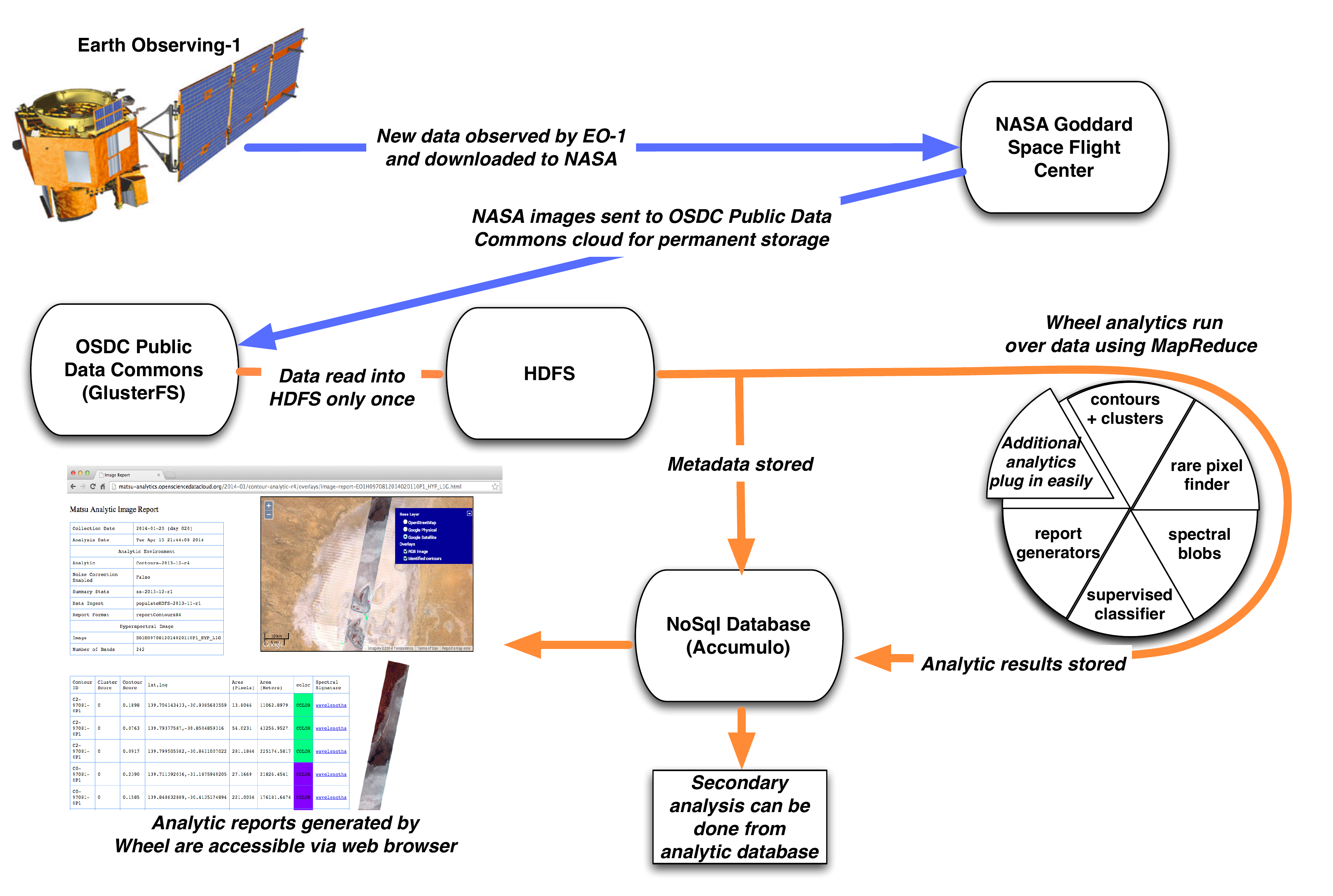}
\caption{A diagram of the flow of EO-1 ALI and Hyperion data
from data acquisition and ingest through the Wheel framework.  Orange denotes
the processes performed on the OSDC Skidmore cloud.  With the Wheel architecture,
the data need to be read in only once regardless of the number of analytics
applied to the data.  The Matsu Wheel system is unique in that, essentially,
the \emph{data} are flowing through the framework while the analytic queries
sit in place and scan for new data.  Additional analytics plug in easily, with the 
requirement that an analytic takes as input a batch of data to be processed.
An analytic query may be written such that it can be run on its own in the Wheel
or written to take as input the output of an upstream analytic.  The report generators
are an example of the latter case, generating summary information from upstream analytics.}
\label{fig:wheel}
\end{figure*}

As many analytics can be included in the Wheel as can run in the allowed time.
If new data are obtained each day, then the limit is 24 hours to avoid back-ups
in processing.  For other use cases, there may be a different time window in which
the results are needed.  This can be seconds, minutes, or hours.  Our MapReduce 
analytic environment is not designed to yield immediate results, but the 
analytics can be run on individual images at a per minute speed.  Analytics 
with results that  need to be made available as soon as possible can be 
configured to run first in the Wheel.  We show a diagram of the flow of EO-1 
data from acquisition and ingest into HDFS through the analytic Wheel framework 
in Figure \ref{fig:wheel}. 

The Wheel architecture is an efficient framework not restricted only to 
image processing, but is applicable to any workflow where an assortment 
of analytics needs to be run on data that require heavy pre-processing or 
have high-volume throughput.

\section{Analytics}
We are currently running five scanning analytics on daily images from 
NASA's EO-1 satellite with the Project Matsu Wheel, including several spectral
anomaly detection algorithms and a land cover classification analytic.  
Here we describe each of these and the resulting analytic reports generated.

\subsection{Contours and Clusters}
This analytic looks for contours in geographic space around clusters in spectral 
space.  
The input data consist of Level 1G EO-1 GeoTiff images from Hyperion, essentially
a set of radiances for all spectral bands for each pixel in an image.  The
radiance in each band is divided by its underlying solar irradiance to convert to 
units of reflectivity or at-sensor reflectance.  This is done
by scaling each band individually by the irradiance and then applying a
geometric correction for the solar elevation and Earth-Sun distance, as shown in
eqn.~\ref{eq:correction}, 

\begin{equation} \label{eq:correction}
\rho_i = \left(\frac{\pi}{\mu_0 F_{0,i} / d_{earth-sun}^2} \right) L_i
\end{equation}
where $\rho_i$ is the at-sensor reflectance at channel $i$, $\mu_0 =
\cos{\textrm{(solar zenith angle)}}$, $F_{0,i}$ is the incident solar flux at
channel $i$, $d_{Earth-Sun}$ is the Earth-Sun distance, and $L_i$ is the
irradiance recorded at channel $i$ \cite{Griffin2005}.  
This correction accounts for differences in the data due to time of day or year. 

We then apply a principal component analysis (PCA) to the set
of reflectivities, and the top N (we choose N=5) 
PCA components are extracted for further analysis.
There are two passes for the Contours and Clusters analytic to generate a 
contour result: spectral and spatial.
\begin{itemize}
  \item Spectral clusters are found in the transformed N-dimensional spectral space 
  for each image using a k-means clustering algorithm and are then
  ranked from most to least extreme using the 
  Mahalanobis distance of the cluster from the spectral center of the image.
  
  \item Pixels are spatially analyzed and are grouped together into contiguous 
  regions with a certain minimum purity or fraction of pixels that belong to that
  cluster and are then ranked again based on their distance from the spectral 
  cluster center.
\end{itemize}

Each cluster then has a score indicating 1) how anomalous the spectral signature is 
in comparison with the rest of the image
and 2) how close the pixels within the contour are to the cluster signature.  
The top ten most anomalous clusters over a given timeframe are singled out for manual 
review and highlighted in a daily overview summary report.

The analytic returns the clusters as contours of geographic regions of spectral
"anomalies" which can then be
viewed as polygonal overlays on a map.  The Matsu Wheel produces image reports 
for each image, which contain an interactive map with options for an 
OpenStreetMap, Google Physical, or Google Satellite base layer and an RGB image 
created from the hyperspectral data and identified polygon contours as options
for overlays. Researchers or anyone interested in the results can view the image
reports online through a web browser.  %

\begin{figure*}[!h ] 
\centering
\includegraphics[scale = .45]{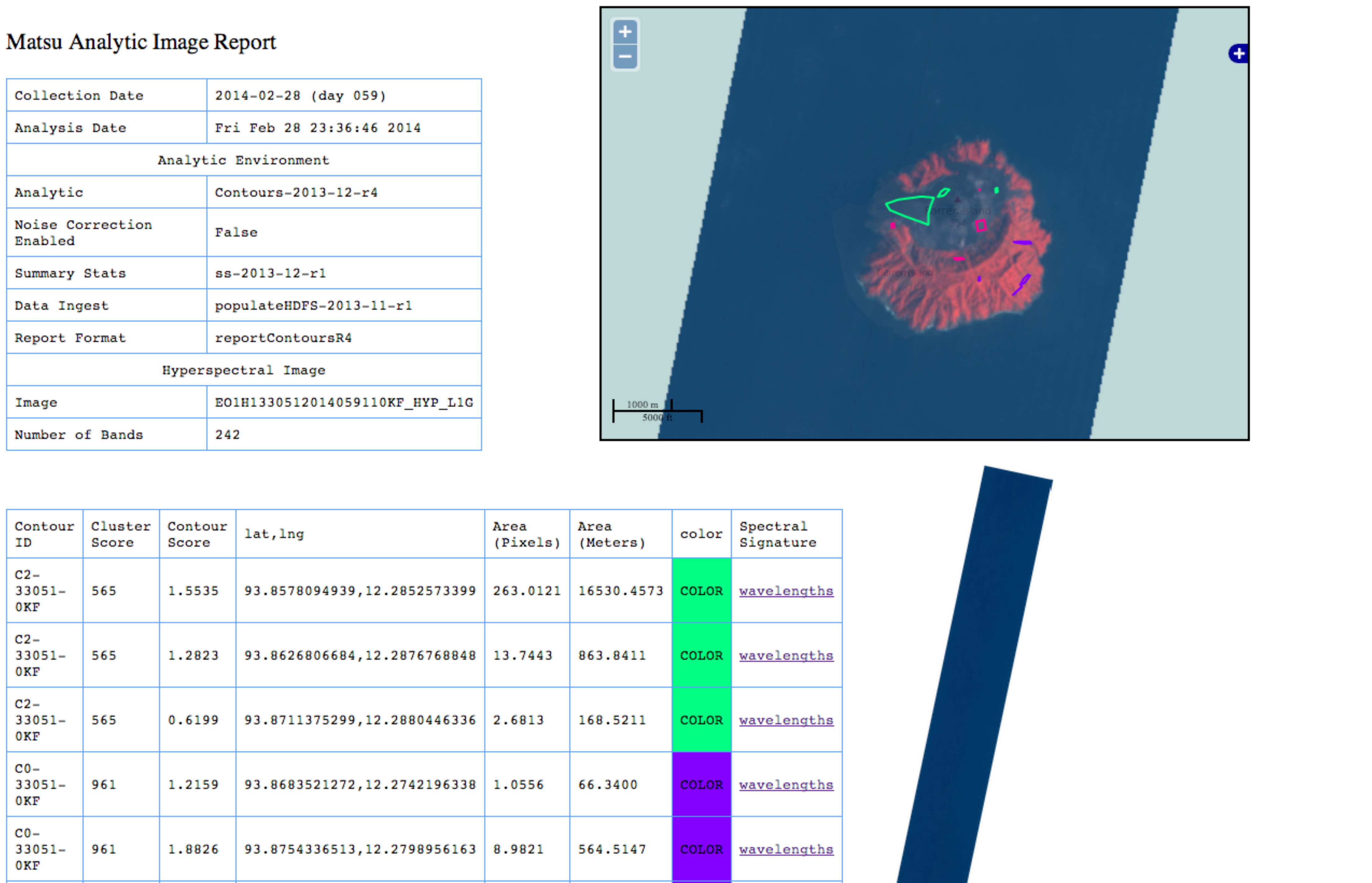}
\caption{A screenshot of a Matsu analytic image report for 
a Contours and Clusters spectral anomaly analytic that automatically identified 
regions of interest
around the Barren Island volcano, confirmed as active by other sources
in February 2014.  The reports contain basic information about the data analyzed
and the analytic products and
a zoomable map with the data and analytic products 
shown as an overlay on an OpenStreetMap, Google Physical,
or Google Satellite base layer.  In this analytic, interesting regions are
given a cluster score from 0 - 1000 based on how anomalous they are compared to the
average detection and appear as colored contours over the image.}
\label{fig:Matsu image report}
\end{figure*}

By implementing this analytic in the Matsu Wheel, we have been able to automatically
identify regions
of interesting activity on the Earth's surface, including several volcanic events.
For example, in February 2014,
this Matsu Wheel analytic automatically identified anomalous activity 
in EO-1 Hyperion data of the Barren Island volcano, which was also confirmed to 
be active by other sources.  We show an example analytic image report for this
event in Figure \ref{fig:Matsu image report}.

\subsection{Rare Pixel Finder}
The Rare Pixel Finder (RPF) is an analytic designed to find small clusters of 
unusual pixels in a hyperspectral image. 
This algorithm is applied directly to the EO-1 data in radiances, but the data can 
also be transformed to reflectances or other metrics or can have logs applied.

Using the subset of k hyperspectral bands that are determined to be most 
informative, it computes k-dimensional Mahalanobis distances and finds the pixels 
most distant.
From this subset, pixels that are both spectrally similar and geographically
proximate are retained. Spectrally similar pixels that can be further grouped into 
small compact sets are reported as potential outlying clusters.
Details of the different steps of the algorithm are given below. 

In the pre-processing step, we remove areas of the image that are obvious anomalies
not related to the image (e.g., zero radiances on the edges of images), as well as spectral
bands that correspond to water absorption or other phenomena that result in near-zero 
observed radiances. Any transformations are applied to the data at this point, such as 
transforming radiance to reflectance or logs.

Once the data are pre-processed, the Mahalanobis distance ($D_i$) is calculated for each pixel.
Then, only the subset $S_1$ of pixels that satisfy $D_i > k_1$ are selected, where $k_1$ is chosen 
such that $S_1$ only contains 0.1--0.5\% of pixels. In practice $k_1$ was based on the upper 
6$\sigma$ of the distribution of sample distances, assuming a log-normal distribution for the
distances. 

For the subset of $S_1$ pixels chosen in the previous step, we next compute a 
similarity matrix $T$ with elements $T_{ij}$ measuring the spectral similarity 
between each pair of pixels. The similarity metric is based on the dot product 
between each pair of points and measures the multi-dimensional angle between points.
The matrix is only formed for the subset $S_1$ and contains a few hundred rows. The pixels
are then further subsetted. Pixel $i$ is selected for set $S_2$ if $T_{ij} > k_2$ for 
$j \neq i$. The parameter $k_2$ is chosen to be very close to 1; in this way we select
objects that are both spectrally extreme but still have similar spectral profiles to one
another.

In order to cluster the pixels geographically, we assume they lie on a rectangular grid. 
An $L_1$ norm metric is applied to the subset $S_2$ in order to further whittle down 
the candidate set to pixels that are spectrally extreme, spectrally similar, and geographically 
proximate. The metric is set so that traversing from one pixel to an immediately adjacent one would 
give a distance 
of $1$, as if only vertical and horizontal movements were allowed. Pixel $i$ is selected for set 
$S_3$ if $M_{ij} < k_3$ for $j \neq i$ and for some
small value of $k_3$. We used $k_3 = 3$, so that pixels either needed to be touching on a face or 
diagonally adjacent.

In the final step a simple heuristic designed to find the connected components of an 
undirected graph is applied. We further restrict to a set $S_4$ that are geographically
proximate to at least $k_4$ other pixels (including itself). We used $k_4=5$, with the
goal of finding compact clusters of between 5 and 20 pixels in size. The clumping
heuristic then returns the pixels that were mapped into a cluster, the corresponding
cluster ID's, and the associated distances for the elements of the cluster.

The flagged pixels are reported as "objects". These groups of objects are further filtered in order
to make sure that they actually represent regions of interest. The following criteria must be 
satisfied in order to qualify:

\begin{enumerate}
  \item {\it Spectral Extremeness}.  The mean Mahalanobis Distance must be greater 
  than or equal to some parameter $p_1$.  This selects clusters that are 
  sufficiently extreme in spectral space.   
  \item {\it Spectral Closeness}.  The Signal-to-Noise ratio must be greater than 
  or equal to some parameter $p_2$.  This selects clusters that have similar 
  values of Mahalanobis Distance. 
  \item {\it Cluster Size}.  All clusters must be between a minimum value 
  parameter $p_3$ and a maximum parameter $p_4$ pixels in size (the goal of this 
  classifier is to find small clusters).  
  \item {\it Cluster Dimension}.  All clusters must have at least some
  parameter  $p_5$ rows and columns (but are not restricted to be rectangles).
\end{enumerate}

Parameters $p_1$ - $p_5$ are all tuning parameters that can be set in the algorithm
to achieve the desired results.

\subsection{Gaussian Mixture Model and K-Nearest Neighbors (GMM-KNN) Algorithm}
The Gaussian Mixture Model and  K-Nearest Neighbors (GMM-KNN) algorithm is also
designed to find small clusters of unusual pixels in a multispectral image.
Using 10--30 spectral bands, it fits the most common spectral shapes to a 
Gaussian Mixture Model (smoothly varying, but makes strong assumptions about 
tails of the distribution) and also a K-Nearest Neighbor model (more detailed 
description of tails, but granular), then searches for pixels that are far from 
common.  Once a set of candidate pixels have been found, they are expanded and 
merged into ``clumps'' using a flood-fill algorithm.  Six characteristics of 
these clumps are used to further reduce the number of candidates.

In short, the GMM-KNN algorithm consists of the following steps.

\begin{enumerate}
  \item Preprocessing: Take the logarithm of the radiance value of each band and 
    project the result onto a color-only basis to remove variations in intensity,
    which tend to be transient while variations in color are more indicative of
    ground objects.
  \item Gaussian Mixture Model: Fit $k$ = 20 Gaussian components to the spectra of all 
    pixels, sufficiently large to cover the major structures in a typical image.
  \item Flood-fill: Expand GMM outliers to enclose any surrounding region that is 
    also anomalous, and merge GMM outliers if they are in the same clump.
  \item Characterizing clumps: Derive a detailed suite of features to quantify 
    each clump, including KNN, edge detection, and the distribution of pixel 
    spectra in the clump.
  \item Optimizing selection: Choose the most unusual candidates based on their 
    features.
\end{enumerate}

\subsection{Spectral Blobs}
This algorithm uses a ``windowing'' technique to create a boundary mask from the
standard deviation of neighboring pixels.  A catalog of spectrally similar blobs,
that contain varying numbers of pixels, is created.
This analytic consists of the following steps:
\begin{itemize}
  \item Label spatially connected components of the mask using an undirected graph
  search algorithm.
  \item  Apply statistical significance tests (t-test, chi-squared) to the
  spectral features of the connected component (the ``blobs'').
  \item Merge spectrally similar regions.
\end{itemize}

The anomalous regions are the small spatial blobs that are not a member of any
larger spatial cluster.

\subsection{Supervised Spectral Classifier}
The Supervised Spectral Classifier is a land coverage classification algorithm
for the Matsu Wheel.  We are particularly interested in developing these analytics
for the detection of water and constructing flood maps to complement the onboard
EO-1 flood detection system \cite{Ip2006}.
This analytic is written to take ALI or Hyperion Level 1G
data and classify each pixel in an image as a member of a given class
in a provided training set.  We currently implement this analytic with a 
simple land coverage classification training set with four possible classes:
\begin{itemize}
\item{Clouds}
\item{Water}
\item{Desert / dry land}
\item{Vegetation}
\end{itemize}

The classifier relies on a support vector machine (SVM) algorithm, using the 
reflectance values of each pixel as the characterizing vector. 
In this implementation of the classifier, we bin Hyperion data to
resemble ALI spectra for ease of use and computation speed in training the 
classifier.  The result of this analytic is a GeoTiff showing the classification
at each pixel.

\subsubsection{Building a training data set}
We constructed a training data set of classified spectra from sections of EO-1 
Hyperion images over areas 
with known land coverage and cloud coverage and confirmed our selections by visual 
inspection of three-color (RGB) images created of the training images.
We used a combinination of Hyperion bands B16 (508.22 nm), B23 (579.45 nm), and 
B29 (640.5 nm) to construct the RGB images.
For each image contributing to the training data set, we only included spectra for
collections of pixels that were visually confirmed as exclusively desert, water, 
clouds, or vegetation. 

The training data set consists of approximately 6,000 to
9,000 objects for each class.
We include spectra for a variety of different regions on the
Earth observed during different times of the year and a range of solar elevation 
angles.  Because absolute values are necessary to directly compare the training 
data to all test images, the raw
irradiance values from the Level 1G data must first be converted to at-sensor 
reflectance using eqn.~\ref{eq:correction}. 
Table \ref{tab:trainingset} lists general properties of the Hyperion scenes 
used in constructing the training set, where the class column indicates which
class(es) (C = cloud, W = water, V = vegetation, and D = desert) 
that scene contributed to.

\begin{table}[!h]
\caption{Scenes included in training set}
\label{tab:trainingset}
\centering
\begin{tabular}{|l|c|r|r|r|}
\hline

Region name& Class & Obs Date & Sun Azim. (\degree) & Sun Elev. (\degree)\\ \hline
Aira & W & 4/18/14 & 119.12 & 49.1 \\ 
San Rossore & C/W & 1/29/14 & 145.5 & 20.9 \\ 
San Rossore & C/V & 8/10/12 & 135.8 & 54.5 \\ 
Barton Bendish & C & 8/22/13 & 142.7 & 43.5 \\ 
Jasper Ridge & V/D & 9/17/13 & 140.7 & 46.9 \\ 
Jasper Ridge & V/C & 9/14/13 & 132.9 & 45.2 \\ %
Jasper Ridge & V/C & 9/27/12 & 147.6 & 45.4 \\ %
Arabian Desert & D & 12/30/12 & 147.0 & 29.9 \\
Jornada & D & 12/10/12 & 28.7 & 151.4 \\
Jornada & D & 7/24/12 & 59.6 & 107.5 \\
Negev & D & 9/15/12 & 130.7 & 52.1 \\
White Sands & C & 7/29/12 & 58.3 & 108.4 \\
Besetsutzuyu & V & 7/14/12 & 56.8 & 135.5 \\
Kenatedo & W & 6/22/12 & 50.5 & 46.5 \\ 
Santarem & W & 6/17/12 & 57.1 & 118.15 \\
Bibubemuku & W & 5/20/12 & 57.5 & 127.7 \\
\hline
\end{tabular}
\end{table}

\begin{figure*}[!h ]
\centering
\includegraphics[scale=0.5]{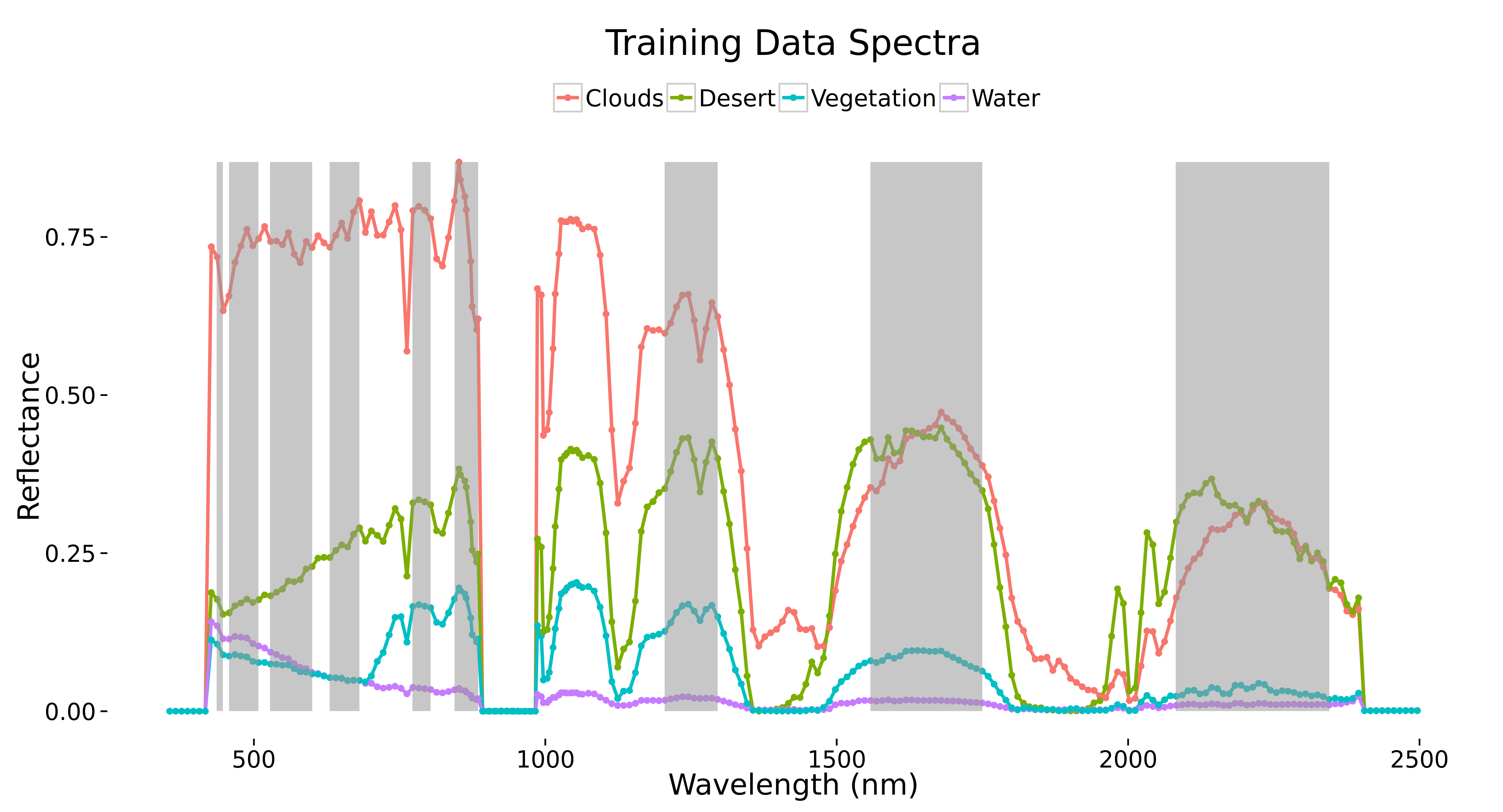}
\caption{The average reflectance spectra for each of the four classifications in
the training data used by our implementation of the Supervised Spectral Classifier
analytic. The four classes are clouds (salmon), desert (lime green), vegetation
(cyan), and water (purple). Shaded grey areas show the wavelength coverage of
ALI bands, which are the wavelength regions used by the classifier described.}
\label{fig:trainingSpec}
\end{figure*}

We show a plot of the average training set object's reflectance spectra for each
of the four classes (clouds, desert, vegetation, water) in Figure 
\ref{fig:trainingSpec}, which are in agreement with the expected results.
These spectra are consistent with the spectral signatures of cloud, vegetation,
and desert sand presented in other examples of EO-1 Hyperion data analysis,
specifically Figures 3 and 4 in Griffin, et. al. (2005)
\cite{Griffin2005} \cite{Burke2004}.

\subsubsection{Classifying new data}
The classifier uses the SVM method provided by the Python
scikit-learn machine learning package \cite{scikit-learn}. 
We construct a vector space from all ALI bands and two additional ALI band ratios,
the ratios between ALI bands 3:7 and 4:8.  These ratios were chosen because
they provided the best individual results in correctly distinguishing between 
classes when used as the sole dimension for the SVM.

For Hyperion images, bands that
correspond to the coverages of the ALI bands are combined. 
The same corrections to reflectance values that are applied to the training data 
are applied to the input image data. For ALI data, an additional scale and offset
need to be applied before the irradiance values are converted to reflectance.

\subsubsection{Validating results}
\begin{figure*}
\centering
	
	\begin{subfigure}[b]{.2\linewidth}
		\includegraphics[scale=0.65]{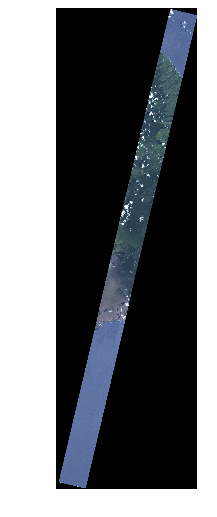}
		\caption{EarthExplorer: \\10-19\% cloud coverage}
		\label{fig:10rgb}
	\end{subfigure}
	\begin{subfigure}[b]{.2\linewidth}
		\includegraphics[scale=0.65]{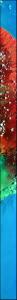}
		\caption{This work: \\0.5\% cloud coverage}
		\label{fig:10class}
	\end{subfigure}
    \hspace{.65in}
    \begin{subfigure}[b]{.2\linewidth}
		\includegraphics[scale=0.65]{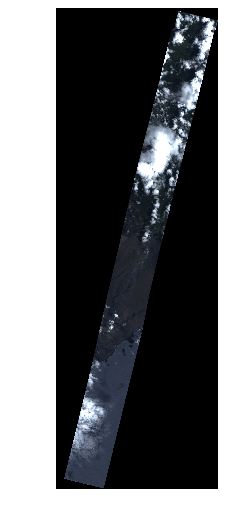}
		\caption{EarthExplorer: \\40-49\% cloud coverage}
		\label{fig:40rgb}
	\end{subfigure}
	\begin{subfigure}[b]{.2\linewidth}
		\includegraphics[scale=0.65]{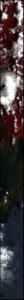}
		\caption{This work: \\23\% cloud coverage}
		\label{fig:40class}
	\end{subfigure}
	
	\begin{subfigure}[b]{.2\linewidth}
		\includegraphics[scale=0.675]{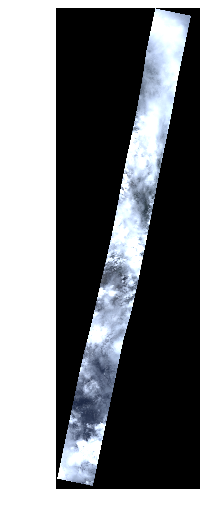}
		\caption{EarthExplorer: \\90-99\% cloud coverage}
		\label{fig:90rgb}
	\end{subfigure}
    \begin{subfigure}[b]{.2\linewidth}
		\includegraphics[scale=0.675]{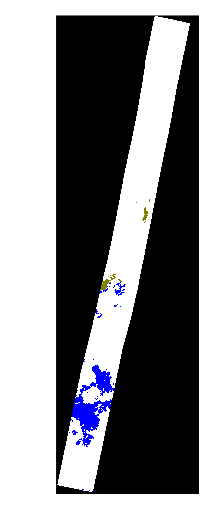}
		\caption{This work: \\95\% cloud coverage}
		\label{fig:90class}
	\end{subfigure}
    \hspace{.65in}
    \begin{subfigure}[b]{.2\linewidth}
        \includegraphics[scale=0.675]{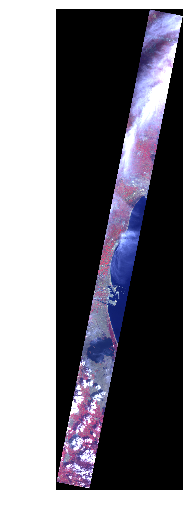}
        \caption{EarthExplorer: \\10-19\% cloud coverage}
        \label{fig:valencrgb}
     \end{subfigure}
     \begin{subfigure}[b]{.2\linewidth}
    	\includegraphics[scale=0.675]{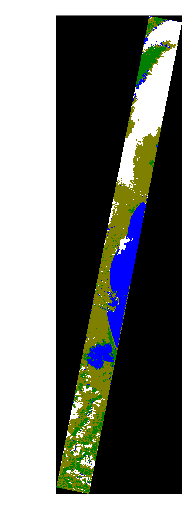}
        \caption{This work: \\28\% cloud coverage}
        \label{fig:valencclass}
    \end{subfigure}
	
	\caption{A visual comparison of RGB images  (left) 
    for several Hyperion test data scenes
    against the results from our Supervised Spectral Classifier  (right) with a 
    training set of 4 classes. For the classified images, white = clouds, 
    green = vegetation, blue = water, brown = desert/ dry land.
    Subfigures a) to f) are scenes over the big island of Hawaii showing a range
    of cloud coverage amounts.  Subfigures g) and h) show the coastal region of
    Valencia, Spain with a good mix of all four classes.
    The classifier results generally appear to be visually consistent with RGB 
    images, though the calculations for classified cloud coverage are not in 
    complete agreement with USGS. This may be because of a different treatment of
    very thin cloud "haze" in cloud coverage calculations.	}
    \label{fig:cloudtests}
\end{figure*}

To confirm that the classifier analytic is generating reasonable results,
we compare the fractional amount of land coverage types calculated by 
the classifier with known fractional amounts from other sources.
We compare our results for classified cloud coverage with the cloud coverage
amounts stated for individual scenes available through the
EarthExplorer tool from the U.S. Geological Survey (USGS) \cite{EarthExplorer}.
We show a visual comparison of cloud coverage determined by our classifier with cloud 
coverage amounts stated by EarthExplorer for three Hyperion scenes of the big island 
of Hawaii with varying amounts of clouds and a randomly chosen scene of a section of 
the coast of Valencia in Figure \ref{fig:cloudtests}. 
For each pair of scenes, the left image shows an RGB image
of the Hyperion data with the
USGS calculated cloud coverage, and the right column has the classified image results,
with the amount of cloud coverage classified indicated. 

The classified images appear to be visually consistent with RGB images, though the
calculations for classified cloud coverage are not in complete agreement
with USGS particularly for lower amounts of cloud coverage. 
This may be because the USGS EarthExplorer images include very thin cloud "haze" 
in cloud coverage calculations.

\begin{figure}
	\includegraphics[scale=0.5]{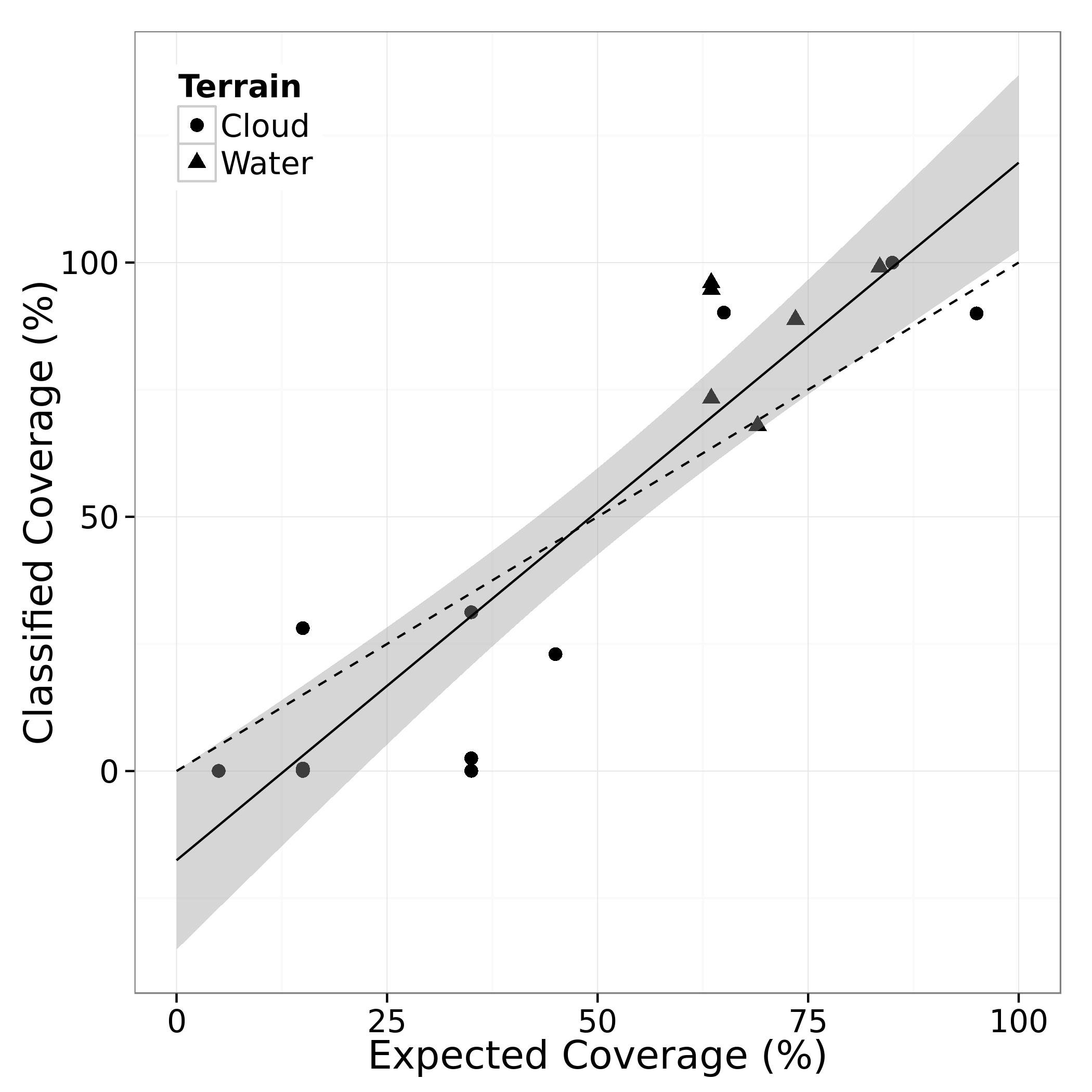}
    \caption{Comparison of expected cloud and water coverages from USGS vs. 
    coverages calculated from our classifier. Expected water points (triangles) are 
    calculated from island scenes as described in the text. 
 	Expected cloud coverage estimates (circles) are taken from USGS EarthExplorer 
    quoted cloud coverage for each image. 
    The linear regression is the solid black line, and the grey 
    shaded area is the 95\% confidence interval. 
    A 1-1 relationship is shown as a dashed black line for comparison.}
    \label{fig:expected}
\end{figure}

In the image of Valencia, Spain, shown in Figures \ref{fig:valencrgb} and
\ref{fig:valencclass} the classifier can clearly distinguish the 
coastline in the picture, correctly classifying the water and land features
and shows good agreement with regions that appear to be cloud.
It has some difficulty in shadowed areas like those regions covered by cloud shadow.
In Figure \ref{fig:expected}, we show a plot comparing expected cloud and water 
coverage to the coverages determined by our classifier for 20 random test scenes. 
For each scene, the expected cloud coverage is taken as the center of the range 
provided by the USGS EarthExplorer summary for that image. The expected water coverage 
is calculated from scenes of islands that are completely contained within the image. 
We can then calculate expected water coverage by removing the known fractional
land area of the islands and the USGS reported cloud coverage.
We fit a regression line to the data, which shows an overall consistent 
relationship between the classified results and expected estimates.

\subsection{Viewing analytic results}
For convenience, each analytic produces a report after each run of the Wheel.  
These reports are built from the JSON results stored in Accumulo 
and are accessible to the public via a web page.  The generated reports
contain spectral and geospatial information about the scene analyzed
as well as analytic results.  An overview summary report is created for 
all daily data processed by an analytic in one run of the Wheel in addition to
reports for individual scenes.  These reports are generated and viewable
immediately upon completion of the scan of new data available each day
at the following address: \url{http://matsu-analytics.opensciencedatacloud.org/}.
Analytic products are also made programmatically accessible through a Web Map
Service.

\section{Future work}
The Matsu Wheel allows for additional analytics to be easily
slotted in with no change to the existing framework so that we
can continue to develop a variety of scanning analytics over these data.
We are extending our existing Supervised Spectral Classifier to use 
specifically over floodplain regions to aid in flood detection for disaster relief.
We are also planning to develop a similar analytic to aid in the detection of fires.

The analytics we described here are all detection algorithms, but we can
also apply this framework and the results of our current analytics to implement
algorithms for prediction. For example, our future work includes developing 
Wheel analytics for the prediction of floods.  This could be done using the following
approach:
\begin{enumerate}
\item Develop a dataset of features describing the observed topology of the Earth.
\item Use the topological data to identify "flood basins," or regions that may
		accumulate water around a local minimum.
\item Determine the relationship between detected water coverage in flood basins and 
		the volume of water present.
\item Use observed water coverage on specific dates to relate the water volume
		in flood basins with time.
\item Use geospatial climate data to relate recent rainfall amounts with water
		volume, which then provides a simple model relating rainfall to expected
        water coverage at any pixel.
\end{enumerate}

This proposed scanning analytic would provide important information 
particularly if implemented
over satellite data with global and frequent coverage, such as data from the 
Global Precipitation Measurement (GPM) mission \cite{GPM} \cite{Neeck2013}.
Our future work involves continuing to develop the Matsu Wheel analytics 
and apply this framework to additional Earth satellite datasets.

\section{Summary}
We have described here the Project Matsu Wheel, which is what we believe to be
the first working application of a Hadoop-based framework for creating analysis
products from a daily scan of available satellite imagery data.  This system is
unique in that it allows for new analytics to be dropped into a daily process
that scans all available data and produces new data analysis products.
With an analytic Wheel scanning framework, the data
need to be read in only once, regardless of the number or types of analytics
applied, which is particularly advantageous when large volumes of data,
such as those produced by Earth satellite observations, need to be 
processed by an assortment of analytics.

We currently use the Matsu Wheel to process daily spectral data from
NASA's EO-1 satellite and make the data and Wheel analytic products available 
to the public through the Open Science Data Cloud and via analytic 
reports on the web. 

A driving goal of Project Matsu is to develop open source technology
for satellite imagery analysis and data mining analytics to provide
data products in support of human assisted disaster relief. The open 
nature of this project and its implementation over commodity hardware
encourages the development and growth of a community of contributors
to develop new scanning analytics for these and other Earth satellite data.

\section*{Acknowledgment}

Project Matsu is an Open Commons Consortium (OCC)-sponsored project
supported by the Open Science Data Cloud.
The source code and documentation is made available on GitHub at
(\url{https://github.com/LabAdvComp/matsu-project}).
This work was supported in part by grants from Gordon and Betty Moore 
Foundation and the National Science Foundation (Grant OISE- 1129076
and CISE 1127316).

The Earth Observing-1 satellite image is courtesy of 
the Earth Observing-1 project team at NASA Goddard Space Flight Center.
The EarthExplorer cloud coverage 
calculations are available from the U.S. Geological Survey on earthexplorer.usgs.gov.

\bibliography{biblio}
\bibliographystyle{IEEEtran}

\end{document}